 \documentclass[reprint]{JASA}

\usepackage{algpseudocode}

\begin{document}

\title[Feng et al.]{I-vector Based Within Speaker Voice Quality Identification on connected speech}
\author{Chuyao Feng}
\affiliation{School of Electrical and Computer Engineering,  Georgia Institute of Technology, Atlanta, GA, 30332, USA}
\author{Eva van Leer}
\affiliation{Department of Communication Sciences and Disorders, Georgia State University, Atlanta, GA,30303, USA}
\author{Mackenzie Lee Curtis}
\affiliation{Department of Communication Sciences and Disorders, Georgia State University, Atlanta, GA,30303, USA}
\author{David V. Anderson}
\affiliation{School of Electrical and Computer Engineering,  Georgia Institute of Technology, Atlanta, GA, 30332, USA}

\preprint{Author, JASA}		

\date{\today} 

\begin{abstract}
Voice disorders affect a large portion of the population, especially heavy voice users such as teachers or call-center workers. Most voice disorders can be treated effectively with behavioral voice therapy, which teaches patients to replace problematic, habituated voice production mechanics with optimal voice production technique(s), yielding improved voice quality. However, treatment often fails because patients have difficulty differentiating their habitual voice from the target technique independently, when clinician feedback is unavailable between therapy sessions. Therefore, with the long term aim to extend clinician feedback to extra-clinical settings, we built two systems that automatically differentiate various voice qualities produced by the same individual. We hypothesized that 1) a system based on i-vectors could classify these qualities as if they represent different speakers and 2) such a system would outperform one based on traditional voice signal processing algorithms. Training recordings were provided by thirteen amateur actors, each producing 5 perceptually different voice qualities in connected speech: normal, breathy, fry, twang, and hyponasal. As hypothesized, the i-vector system outperformed the acoustic measure system in classification accuracy (i.e. 97.5\% compared to 77.2\%, respectively). Findings are expected because the i-vector system maps features to an integrated space which better represents each voice quality than the 22-feature space of the baseline system. Therefore, an i-vector based system has potential for clinical application in voice therapy and voice training. 
\end{abstract}

\maketitle

\section{\label{sec:intro} Introduction}

Voice disorders are changes in voice quality, pitch, loudness, effort and endurance \cite{ramig1998treatment}. These affect approximately 7.5 million Americans at a time and can occur throughout the lifespan, with multiple etiologies and a prevalence of up to 47\%in  teachers \cite{thibeault2004occupational,cohen2012impact,cohen2010self}. Most voice disorders can be treated effectively with behavioral voice therapy provided by speech-language pathologists (SLPs) \cite{schwartz2009clinical} \cite{ramig1998treatment}.However, as with other behavioral interventions \cite{glanz2008health} (e.g., diet, exercise, smoking cessation), patient adherence to voice therapy is often poor \citep{smith2010patient,van2012use} and drop-out rates that exceed 50\% \cite{portone2008review,hapner2009study}.This problem largely stems from patients' difficulty in altering their voice use outside of therapy session, when they do not have access to clinician feedback.

In weekly voice therapy sessions, the speech language pathologist teaches the patient to replace their habitual, sub-optimal voice production mechanics with an improved voice production technique such as resonant, breathy, twang, or loud voice \cite{ramig1998treatment}. This target technique is individualized to the patient and associated with voice quality improvement. For example, patients with a weak, excessively breathy voice due to aging or unilateral vocal fold paresis may be taught to apply techniques of increased vocal effort and twang resonance to reduce breathiness \cite{ziegler2014preliminary,lombard2007novel}. Conversely, patients with a strained voice quality due to adducted hyperfunction (i.e. excessively effortful voice production) might reduce strain by increasing breathiness  \cite{ziegler2014preliminary}. Thus a particular voice quality such as breathiness may be problematic in one patient and a therapeutic target in another. Furthermore, use of the target technique may a alter untargeted vocal parameters such as loudness,  fundamental frequency and speaking rate.\cite{dejonckere2001plasticity,bonilha2012creating}. Thus, the so-called target voice varies from the patient's habitual voice along multiple parameters or features of voice and speech.

While patients may produce the target technique successfully in weekly treatment sessions, they commonly struggle to replicate it between sessions, without the therapist's feedback\cite{van2010patient}.  Specifically, patients report difficulty differentiating their habitual voice from the therapeutic target in dedicated technique practice as well as daily conversation.  Unlike the bulls eye in a game of darts, a vocal target is not visually apparent, making it hard to judge one's progress toward it \cite{van2010patient}. Consequently, patients become demoralized, practice infrequently, and fail to replace their habitual voice with the target voice \citep{}. Even those who succeed in therapy fail to maintain the voice change in the long term, relapsing back to a pre-therapy vocal status \citep{}. 

Since patients have difficulty producing the target voice without clinician feedback they may benefit from automated supplemental feedback systems. At a minimum, such systems could differentiate the patient's habitual voice quality from their  therapeutic target  in running speech. Such feedback would extend the clinician's judgement from the therapy session to the extra-clinical environment. Since each patient's habitual and target voice qualities are unique, the system would aim for  an individualized comparison between the habitual and target production for each patient, rather than a comparison to standardized norms. Furthermore, the system should remain robust in noisy speaking environments. No such system presently exists. 

Extensive research into objective acoustic measures of voice has been conducted in an effort to replace subjective expert judgement of voice quality with objective numbers. Several measures of the voice signal have been associated with expert judgement of breathiness, overall clarity of voice, and to a lesser degree, roughness and strain \citep{} . In order to replace the expert clinician's judgement of voice quality, several systems have been built to classify voice samples. Approaches has included using acoustic features with machine learning methods on sustained vowels. The majority of such studies have aimed to automatically detect features associated with emotion, e.g. such as acoustic signs of depression or anger \citep{scherer2000psychological}. Another group of studies has sought to predict the pathology label (i.e. diagnosis) associated with a database of voice samples, without examining voice quality. A handful of studies have aimed to classify voice qualities across speakers. Mehta group developed a system that differentiates modal (i.e. normal) voice quality from breathy, rough and strained production on sustained vowels \cite{mehta2012mobile}. In this study, 28 actors provided samples of modal, breathy, strained and rough quality. Actors variability in producing the targets was a limitation to accuracy classification, since multiple recordings did not fully or consistently reflect the target voice quality. Wang automatically classified breathy and rough voice ? from normal voice in sustained phonation recordings of disordered voice labeled perceptually judged with the GRB parameters of the GRBAS with best accuracy around 80\% \citep{wang2016automatic}.  Similarly, \cite{arias2019multimodal} used deep learning framework with RGB scale and MFCC features, which shows an improvement of accuracy of 18.1\%. In sum, past approaches have focused largely on automatically detecting specific voice qualities in sustained phonation but not on connected speech.
   
Objective acoustic measures of the voice signal have been developed to supplement or replace human perceptual judgment for research and outcome purposes, but few have been examined as extra-clinical patient feedback. Mobile feedback regarding vocal loudness and pitch can be provided through a variety of existing technologies and apps \citep{}. However, objective electronic measurement of voice quality (i.e., timbre) has proven more challenging, in turn making provision of voice quality feedback more difficult. Cepstral peak prominence (CPP), which quantifies aspects of signal periodicity and harmonic energy, is significantly and  inversely correlation with perceived dysphonia, such that it may be used a measure of overall voice quality. CPP has demonstrated utility as feedback for patients practicing a resonant voice target \citep{}.  However, results of CPP and other voice measures require interpretation in order to determine if the patient is speaking in his or her problematic, habitual voice quality or their prescribed target voice quality (i.e., good voice). Furthermore, depending on the patient and the target, additional parameters may be needed to consistently differentiate a patient's habitual and target voice, yielding more measures for the patient to interpret. Most recently, algorithms comprised of multiple weighted acoustic measures have been developed to best correlate with human perception of overall vocal clarity or overall degree of dysphonia \cite{van2017ios}, but it cannot be applied to connected speech and have not been tested for utility in providing voice quality feedback to patients. A more comprehensive feedback system is under development to provide information regarding relevant aspects of voice production mechanics but requires the patient to wear an accelerometer taped to his or her neck \cite{ghassemi2014learning}. In sum, systems that capture vocal clarity are under development for patient use. 

There are two limitations to the current approaches. First, in therapy, qualities that constitute the problematic habitual voice for a certain patient may represent the target voice for another patient, depending on their particular disorders and individual differences. Thus, voice quality goals should be set in reference to within-speaker parameters rather than in reference to between-speaker norms. Second, clinical judgment of voice quality may involve multiple parameters that require knowledge from the treating clinician. Therefore, there is a need for a system that does not replace the clinician’s ability to detect a bad (i.e. habitual) from a good voice  (i.e. target), but rather extends the clinical judgment outside of the treatment room. Lastly, previous approaches have classified sustained phonation samples that do not represent the patient's target voice use in connected speech. 

With the larger aim to develop an automated feedback system for patient use, the objective of the present study was to build and validate a system that can differentiate various within-speaker voice qualities in connected speech. This system would yield a proof of concept that within-speaker voice quality targets can be detected and differentiated automatically in running speech. In the present study, thirteen actors each produced 5 different voice qualities to yield a data set for developing and testing the system: normal (clear), breathy, fry, twang, and hyponasality (i.e. voice quality associated with having a headcold). Achieving automatic differentiation of these quality extremes represents a first step in the development of a system that can differentiate good versus bad voice qualities within the running speech of voice patients. If these grossly differing voices can be differentiated automatically, future work is indicated for detection of more subtle differences 

The aim will be explored through two different methodological approaches. Our first approach is to incorporate common acoustic measures of vocal parameters as ingredients for voice quality differentiation, including Pitch Strength (PS), CPPS, Harmonic-to-Noise Ratio (HNR), and other fundenmental frequency related statistics. Since these measures are commonly used to detect vocal parameters, it is reasonable to expect that their combination may successfully serve as ingredients to differentiate within-speaker voices. It serves as a baseline for our second i-vector based approach for comparison.

Our second method uses i-vector, which works well in speaker recognition and we hypothesize that i-vectors can be applied to identify different intra-speaker voice qualities. In past years, i-vector methods have demonstrated a state-of-the-art performance of text-independent speaker verification systems \cite{dehak2011front}. These were traditionally developed to detect differences between speakers (i.e., speaker identification and verification) but did not detect intraspeaker voice quality differences. Common voice qualities that may constitute either target or habitual voice qualities include vocal fry, breathiness, twang, and hyponasal resonance, and an overall clear (normal) voice quality. Depending on the patient,  some of these qualities may represent bad and good technique. The system effectively examines a speaker’s different vocal qualities as if these represent different speakers (i.e., your good-voice-self vs. your bad-voice-self). The approach also differs conceptually from current patient-centered acoustic analysis in requiring that the patient wear an accelerometer taped to their neck \cite{mehta2012mobile} or seeks to entirely replace the therapist \cite{maryn2010toward} rather than extend the therapist’s judgment (i.e., ``now you are in your good voice!'') to the patient’s environment \cite{ghassemi2014learning}. Once an SLP is involved, these voice quality metrics do not provide additional benefit.  In contrast, the presently presented system allows the therapist to use their judgment to set targets and provides feedback to the patient on whether they are speaking in the target voice. The purpose of this paper is to develop, test and compare two methods for automatic voice quality differentiation within speakers.


This paper is organized as follows: Section~\ref{sec:data} describes the data collection process.Section 1 describes the pipeline of system 1 involving common acoustic measures Section~\ref{sec:ivector} describes the general pipeline of the system 2 based in i-vectorsThe experiment setup and result with clean and noise data is described in Section~\ref{sec:setup} and Section~\ref{sec:result}, followed by the discussion in Section~\ref{sec:discussion} and the conclusion in Section~\ref{sec:conclusion}.

\begin{figure}
  \figline{\fig{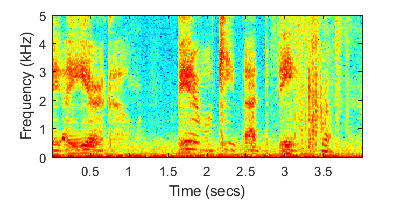}{0.85\linewidth}{(a)}\label{fig:fry>}}
  \vskip-0.1in
  \figline{\fig{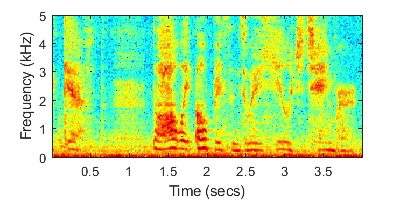}{0.85\linewidth}{(b)}\label{fig:breathy>}}
  \vskip-0.1in
  \figline{\fig{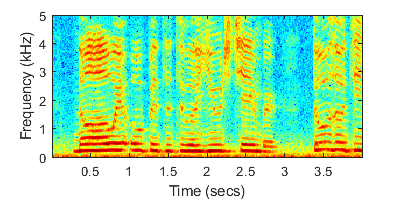}{0.85\linewidth}{(c)}\label{fig:twang>}}
  \vskip-0.1in
  \figline{\fig{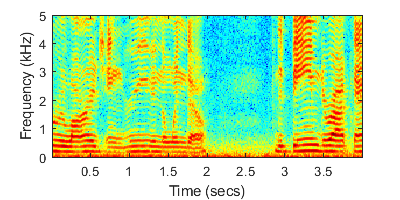}{0.85\linewidth}{(d)}\label{fig:hyponasal>}}
  \vskip-0.1in
  \figline{\fig{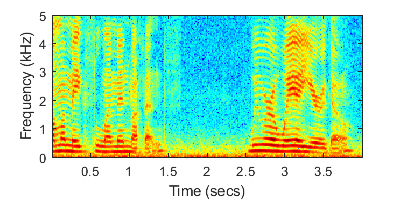}{0.85\linewidth}{(e)}\label{fig:normal>}}
  \vskip-0.1in
  \caption{ \label{fig:spectro} Spetrogram of five different voice quality modes. (a) Vocal fry. (b) Breathiness. (c) Twang (d) Hyponasal resonance. (e) Normal.}
\end{figure}

\section{\label{sec:data}Data Collection}
13 adults (10 women and 3 men) ages 18-49 (mean=26.9, SD=9.7) were enrolled in the study to provide data for algorithm development.\footnote{IRB approval was obtained from Georgia Institute of Technology and Georgia State University.} Participants were screened for normal voice quality by the third author, a licensed speech-language pathologist, and a graduate research assistant. Additionally, participants summarized their voice as ``normal'' (i.e {absence of impairment) on the Voice Handicap Index (VHI) \cite{jacobson1997voice}. The average VHI score was 12.4 (SD=16.38) on this self-report measure; this was slightly higher than normal due to personality factors that elevated the score for 2 participants. To provide voice data for algorithm development, participants were instructed to produce four different vocal qualities in addition to their habitual voice: breathy, fry, twang, and hyponasal in spontaneous speaking. The qualities of strain and roughness were excluded from this initial study because actors complained of excessive laryngeal effort when attempting these. Resonant voice was omitted because the learning curve for this target technique was too long in duration. Breathiness is presence of audible air escape in the voice \citep{kempster2009consensus} common to voices such as Marily Monroe. Twang is the use of loud, bright (i.e. "brassy") voice quality common in country western singing  \citep{lombard2007novel} and the character Fran Drescher in the comedy series The Nanny. Fry is phonation at a low frequency register below modal voice \citep{hollien1966nature} common in voice disorders and more prevalent in young women than men \citep{hornibrook2018creaky}, and associated with the Valley Girl sociolect. A hyponasal quality is a denasal use of resonance that sounds like one is speaking with a stuffed up nose \cite{behrman2002effect}.  
 Actors were taught to use each voice quality via modeling, direct instruction, and cueing by the second author. Only the quality "hyponasal" was difficult for a subset of speakers, who were allowed to imitate the cartoon character Barney Rubble instead. The resulting S Voice quality label was verified by a research assistant trained to detect each in a forced choice. Once these were elicited successfully, uncompressed audio recording was performed at a sampling rate of 44.1KHz in a quiet room using the Roland RO-5 wave/mp3 recorder. Participants were asked to read 26 standard sentences out loud, and subsequently produce  4 minutes of extemporaneous speaking in their normal voice quality, followed by the same tasks in the targeted 4 voice qualities. Standard sentences included the tasks of the Consensus Auditory Perceptual Evaluation of Voice (CAPE-V) \cite{kempster2009consensus}, and ten sentences each of the (TIMIT \cite{zue1990speech} and Harvard sentence inventories \cite{rothauser1969ieee}. FIG.~\ref{fig:spectro} shows the spectrogram of the five different voice qualities of the same speaker. Notice that for FIG.~\ref{fig:fry>} and Fig.~\ref{fig:breathy>}, the harmonic in higher frequency appears blurred in the spectrogram. However, for FIG.~\ref{fig:hyponasal>} and FIG.~\ref{fig:normal>}, there are strong harmonics, and voice quality differences are not immediately apparent.}

\begin{figure*}[t]
\centering
\includegraphics[width=6.8in]{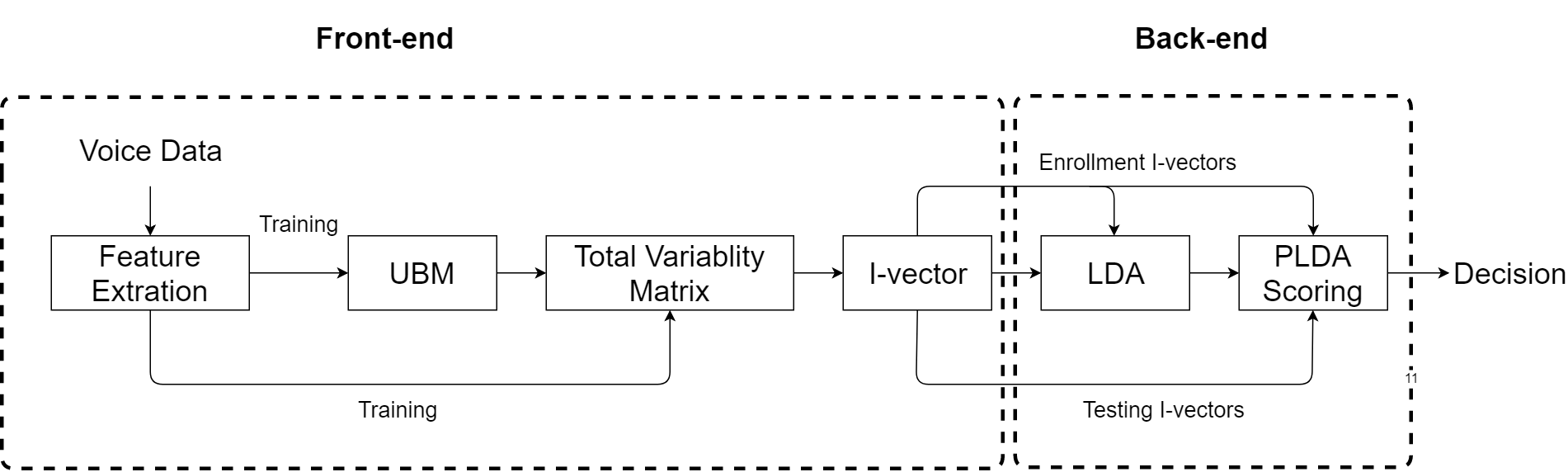}
\caption{Pipeline of the i-vector system}
\label{fig:block}
\end{figure*}

\section{Acoustic Measure System (Baseline)}
There has been a great amount of research searching for acoustic features that could measure the effectiveness of voice therapy treatment outcomes and quantify the severity of dysphonia along a variety of parameters. We explored acoustic measures of voice that have been validated in past research to detect different vocal parameters related to quality and pitch. These were incorporated to create a baseline voice quality detection systemfor our study. . Most of the measures can only be applied to sustained vowels because they may not be accurate on running speech. For example, H1-H2 extracted from vowels has been shown to be effective in measuring breathiness associated with dysphonia \cite{narasimhan2017spectral}. Also, relative fundamental frequency measure have shown to be sensitive to hyperfunction-related voice disorders \cite{stepp2010impact}. Our study focuses on classifying different voice qualities in unscripted speech, so only features that can be applied to connected speech are studied in this paper. We used these acoustic measures with Support Vector Machine to build a voice quality identification that act as a baseline performance of this task.

\subsection{Pitch Strength (PS)}
Pitch strength (PS) has been used widely in sound perception, and recent studies have exhibited that PS may be used to perceptually measure voice quality \cite{rubin2019comparison}. It is an estimate of pitch saliency or pitch tonality, which is usually reported subjectively. PS is different from pitch height, as two notes with the same pitch height could have different pitch strength \cite{kopf2017pitch}. Previous studies have shown that PS has an inverse relationship with breathiness\cite{eddins2016modeling}. Pitch strength is extracted using the Auditory Sawtooth Waveform Inspired Pitch Estimator—Prime (Auditory-SWIPE). The detail of calculating the pitch strength can be found in \cite{camacho2008sawtooth}. The average pitch strength of each recording is used.

\subsection{Cepstral Peak Prominence (CPP)}
Cepstral Peak Prominence (CPP) measures the amount of cepstral energy in a voice sample. It can be applied to both sustained vowel and connected speech\cite{rubin2019comparison,heman2002relationship}.  Similar to the MFCC, the audio signal is initially converted to spectrum. Then, the spectrum is converted to cepstral domain. The cepstral peak is subsequently located, and the difference between the cepstral peak and the value under this peak is calculated CPP has been shown to have an positive association with perceived clarity of voice, such that higher values represent clearer voice, and lower values are associated with more aperiodic and breathy voice quality. CPP also varies with loudness of the signal, such that higher CPP scores can indicate a louder signal
\cite{hillenbrand1994acoustic}. CPP is also used for calculating othercomposite measures of overall voice quality, including the Cepstral Spectral Index of Dysphonia (CSID) and the Acoustic Voice Quality Index (AVQI)\cite{rubin2019comparison}. The AVQI algorithm is a multi-varia measure that includes the CPPS, harmonics-to-noise ratio (HNR) and other acoustic features such as shimmer. However, it is not applicable to connected speech\cite{lee2018comparison}. Unlike AVQI, CSID can be used to analyze connected speech and sustained vowel separately and has been extensively validated\cite{peterson2013toward}. We used the equation in \cite{lee2018comparison} to calculate CSID. Praat is used to extract both CPP and CSID.

\subsection{Harmonic-to-Noise Ratio (HNR)}
Harmonicity can be used to measure the signal-to-noise ratio (SNR) of anything that generates a periodic signal. It is also a good indicator of different voice qualities\cite{jannetts2014cepstral}. Calculation of harmonic-to noise ratio (HNR) is based on the assumption that the signal has a periodic component and an addictive noise component. HNR calculates the energy ratio of signal and noise components. Praat is used to generate HNR.

\subsection{Fundamental frequency F0 different statistics }
	
Fundamental frequency (F0) is considered the lowest frequency of a periodic signal. It has proven to measure aging in womens' voices \cite{da2011acoustic}. Both elevated and lowered F0 may be associated with a variety of pathologies, as well as with with improved vocal outcomes \cite{da2011acoustic}. The fundamental frequency is extracted using Praat. The mean, standard deviation, maximum, minimum and slope of F0 are/is calculated afterward and used as a combined measure for voice quality. 

\section{I-vector based identification system\label{sec:ivector}}

In this paper, we extracted i-vectors as features to identify intra-speaker voice quality variations using the data described above. The problem is defined as a close set identification problem with 5 variations per speaker with the goal to identify different voices quality within speakers. The i-vector pipeline of the voice quality identification system is a series of generative models. A universal background model (UBM) is used to collect sufficient information to compute Baum-Welch statistics \cite{kenny2012small}. A total variability matrix is trained for i-vector extraction using voice features and a classifier back-end is used for identification. The proposed system is shown in Fig.~\ref{fig:block}.

\subsection{Voice Features}
Mel Frequency Cepstral Coefficients (MFCCs) are used as the voice features, which are pre-designed features intended to capture the characteristic of human speech. They are computed from mean-normalized audio clips. Delta and acceleration of MFCCs are appended to generate a larger feature vector. Cepstral mean normalization (CMVN) is applied to the feature vector to remove any channel effects. For practical application, voice activity detector is not applied to keep silence segments in training. 

\subsection{Universal Background Model}
 For the majority of  speaker identification systems, UBM is usually trained with the maximum amount of data from from publicly available speech database. However, there is no concrete evidence that using the maximum amount of data would guarantee the best overall performance \cite{hasan2011studyubm}. Therefore, we trained a Guassian Mixture Model (GMM) on only the training data with model parameters $\lambda = \{ {\pi_c}, {m_c}, {\Sigma_c} \}$.   Each mixture component $c$ = 1, ... ,$C$ , $\pi_c$, $m_c$, $\Sigma_c$ denotes the weights, mean vectors. and covariance matrices. The dimension of the mean vector $m_c$ is $F$ is the same as the dimension of the feature vectors. The dimension of the covariance matrix is $\Sigma_c$ is $F \times F$. UBM is trained using the  Expectation-Maximization algorithm \cite{kenny2005joint}. 
 
\subsection{I-vector Representation and LDA}
Following the notation in \cite{kenny2005joint}, $\textit{\textbf{m}}$ denotes the supervector concatenating the mean of each mixture component $m_1,...,m_C$ with a dimension of $CF \times 1$. For each different voice $v$ (from  the same or different speakers), $\textbf{\textit{M}}(v)$ denotes the voice-adapted supervector for the same voice. For each mixture component, $M_c(v)$ is the concatenation of $m_c(s)$ for $c = 1...C$, which is a subvector of $\textit{\textbf{M}}(v)$. Then, the i-vector representation of the system is:
\begin{eqnarray}
\textit{\textbf{M}}(v) = \textit{\textbf{m}} + \textit{\textbf{Tw}}(s)
\label{equ:1}
\end{eqnarray}

where $\textit{\textbf{w}}(s)$ denotes the i-vector of dimension $M$, and $\textit{\textbf{T}}$ denotes the total variability matrix, which has a dimension of  $CF \times M$. The total variability maps the high dimension supervector of GMM into $\textit{\textbf{w}}(s)$, which is a lower dimensional representation of different voices. The total variability matrix is trained on all of the training data using the EM method and i-vector is computed using the same training data.

LDA is then performed on i-vectors to maximize inter-class variation and to minimize intra-class variances. This algorithm is widely used for dimension reduction in classification problems. After LDA dimension reduction, the low dimension i-vectors are centered and length normalized \cite{garcia2011analysis} before feeding into the classifier.

\section{\label{sec:setup}Experiments}
\subsection{Data Pre-processing}
13 actors, including 3 males and 10 females were recorded reading aloud and speaking extemporaneously in 5 different voice qualities for approximately 5 minutes per voice. All speakers recorded  in  fry, breathy, normal, hyponasal, and twang voice qualities. For all of the recordings, we cut off the first and last 30 seconds, since they are mostly likely to contain conversations between the SLP and the speakers. Each recording was then segmented into 8 second audio sub-segments, resulting in a total of 30 audio segments of each voice quality for each speaker. Next, files were further divided into eight-second segments, four-second segments and two-second segments to evaluate the performance on short recordings. 

\subsection{Feature Extraction}

For the traditional baseline system, we extracted MFCCs, PS, CPPS, CSID, HNR and F0 statistics. MFCCs are 13 dimension and averaged across time, F0 statistics are 5 dimension and other measures are 1 dimension. In total we obtained a 22 dimension feature set for each audio clip.

For our i-vector identification system, 13 MFCCs were extracted with a frame length of 25ms and sliding length of 10ms. 13 delta and 13 double-delta of MFCCs are concatenated with MFCCs to create 39 dimension spectoral feature vectors. The MFCCs wer extracted using a HTK toolkit \cite{htkyoung2002}

\subsection{Classifiers}
Our experiment used two types of classifiers: PLDA scoring \cite{kenny2010bayesian} and multi-class SVM. The PLDA model was trained using the same training data and implemented using MSR toolkit \cite{sadjadi2013msr}. We simply selected the model with highest score. The multi-class SVM was implemented using the lib-svm library \cite{chang2011libsvm}. We use linear kernel SVM with a complexity component value equal to one.
We use both PLDA scoring and SVM in i-vector based system but only SVM in our baseline system.

\subsection{I-vector System Setup}
 The UBM is a 256 component Gaussian mixture model and the dimension of i-vector is 100, since our dataset is small. We trained the UBM and total variability matrix using the first 20 sub-segments of each speaker. The dimension of the i-vector was reduced to 64 after implementing LDA. We also computed the dimension reduced i-vector for the last ten testing sub-segments. The dimension reduced i-vector is used as the input feature for training the classifier.
 
\section{Experimental Result}
\label{sec:result}

\subsection{Intra-speaker Classification}
\begin{figure}[!hbt]
\centering
\includegraphics[width=3.5in]{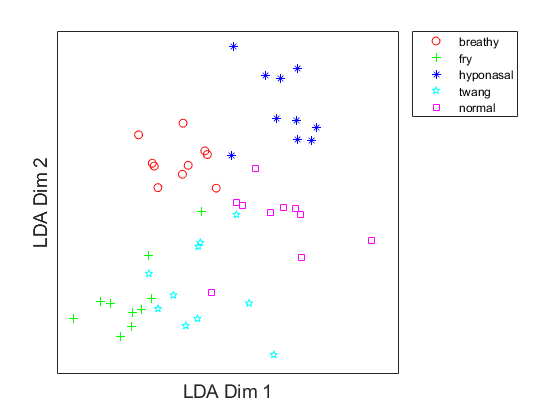}
\caption{First two LDA dimensions of intraspeaker voice qualities}
\label{fig:lda}
\end{figure}

In this experiment, we evaluated the performance of two systems on intra-speaker voice quality variations. For a given testing audio segment, we performed classification only on the five voice models (breathy, fry, twang, hyponasal, and normal) from the same speaker. In the first experiment,we used eight-second audio segments from each actor. However, for practical application, it is important to provide feedback for patients after only minimal delay.  Therefore, we further segmented the testing audio segments to 4 seconds (20 clips for each voice) and 2 seconds (40 segments for each voice) and ran the same experiment on shorter segments to evaluate the feasibility of feedback streaming. TABLE \ref{tab:intraeer} summarizes the performance on intra-speaker voice quality variations of the proposed i-vector system. As seen from the table, both PLDA scoring and multi-class SVM performed well with an average accuracy of 97.5\% and 96.4\% for 8 second audio clips respectively. Our results also demonstrate that using  longer testing utterances work better than a shorter testing utterances. The results meet our expectations, since the percentage of non-speech frames in the shorter audio segment is higher. Although the performance degraded with shorter clips, the average of accuracy remained above 90\%. FIG. \ref{fig:lda} shows the scatter plot of the first two LDA dimension of all the five models of actor nine of our i-vector system. As seen in the plot, the i-vector effectively separates different voice qualities.  

TABLE \ref{tab:comparison} shows the result of our proposed system compared to traditional system. We can see that our system outperforms the traditional system by more than 20\%. The traditional system only achieves 77.2\% on 8-second segments. On 2 second segments, the i-vector system outperforms by more than 30\%.

\begin{table}[!htb]
\renewcommand*{\arraystretch}{1.2}
\caption{Accuracy of intra-speaker experiment of i-vector system.}
\label{tab:intraeer}
\centering
\begin{tabular}{lcccccc}
\hline\hline
          & \multicolumn{3}{c|}{PLDA Scoring}     & \multicolumn{3}{c}{Multi-class SVM} \\ \cline{2-7} 
          & \multicolumn{1}{c}{8s} & \multicolumn{1}{c}{4s} & \multicolumn{1}{c|}{2s} & \multicolumn{1}{c}{8s} & \multicolumn{1}{c}{4s} & \multicolumn{1}{c}{2s}  \\ \cline{2-7}
Actor 1   & 98\%   & 94\%     & 90\%       & 98\%       & 95\%       & 90.5\%     \\
Actor 2   & 100\%  & 99\%     & 97\%       & 96\%       & 93\%       & 91\%       \\
Actor 3   & 98\%   & 97\%     & 93.5\%     & 98\%       & 96\%       & 92.5\%     \\
Actor 4   & 92\%   & 90\%     & 79.5\%     & 78\%       & 77\%       & 70\%       \\
Actor 5   & 100\%  & 96\%     & 90\%       & 96\%       & 92\%       & 85.5\%     \\
Actor 6   & 100\%  & 97\%     & 94.5\%     & 100\%      & 93\%       & 91.5\%     \\
Actor 7   & 100\%  & 100\%    & 96\%       & 100\%      & 99\%       & 96.5\%     \\
Actor 8   & 98\%   & 95\%     & 88.5\%     & 94\%       & 90\%       & 83.5\%     \\
Actor 9   & 100\%  & 100\%    & 99\%       & 100\%      & 100\%      & 99.5\%     \\
Actor 10  & 82\%   & 79\%     & 77.5\%     & 92\%       & 92\%       & 89.5\%     \\
Actor 11  & 100\%  & 100\%    & 99\%       & 100\%      & 100\%      & 98\%       \\
Actor 12  & 100\%  & 97\%     & 91.5\%     & 100\%      & 97\%       & 92\%       \\
Actor 13  & 100\%  & 100\%    & 100\%      & 100\%      & 99\%       & 99\%       \\ 
\cline{2-7}
Average   & \textbf{97.5\%}   & 95.7\%     & 92.0\%     & \textbf{96.4\%}     & 94.1\%     & 90.7\%  \\ \hline\hline
\end{tabular}
\end{table}

\begin{table}[!htb]
\renewcommand*{\arraystretch}{1.2}
\caption{Performance comparison of i-vector system and traditional system}
\label{tab:comparison}
\centering
\begin{tabular}{lcccccc}
\hline\hline
          & \multicolumn{3}{c|}{I-vector System}     & \multicolumn{3}{c}{Acoustic Measure System} \\ \cline{2-7} 
          & \multicolumn{1}{c}{8s} & \multicolumn{1}{c}{4s} & \multicolumn{1}{c|}{2s} & \multicolumn{1}{c}{8s} & \multicolumn{1}{c}{4s} & \multicolumn{1}{c}{2s}  \\ \cline{2-7}
Actor 1   & 98\%   & 94\%     & 90\%       & 86\%       & 77\%       & 70.5\%     \\
Actor 2   & 100\%  & 99\%     & 97\%       & 82\%       & 71\%       & 68\%       \\
Actor 3   & 98\%   & 97\%     & 93.5\%     & 74\%       & 66\%       & 63\%     \\
Actor 4   & 92\%   & 90\%     & 79.5\%     & 68\%       & 58\%       & 53.5\%       \\
Actor 5   & 100\%  & 96\%     & 90\%       & 74\%       & 59\%       & 52\%     \\
Actor 6   & 100\%  & 97\%     & 94.5\%     & 68\%      & 60\%       & 52.5\%     \\
Actor 7   & 100\%  & 100\%    & 96\%       & 76\%      & 63\%       & 51.5\%     \\
Actor 8   & 98\%   & 95\%     & 88.5\%     & 72\%       & 54\%       & 50\%     \\
Actor 9   & 100\%  & 100\%    & 99\%       & 76\%      & 69\%      & 62.5\%     \\
Actor 10  & 82\%   & 79\%     & 77.5\%     & 78\%       & 58\%       & 56\%     \\
Actor 11  & 100\%  & 100\%    & 99\%       & 88\%      & 80\%      & 68.5\%       \\
Actor 12  & 100\%  & 97\%     & 91.5\%     & 80\%      & 71\%       & 61\%       \\
Actor 13  & 100\%  & 100\%    & 100\%      & 82\%      & 73\%       & 60.5\%       \\ 
\cline{2-7}
Average   & \textbf{97.5\%}   & \textbf{95.7\%}     & \textbf{92.0\%}     & 77.2\%     & 66.1\%     & 59.2\%  \\ \hline\hline
\end{tabular}
\end{table}

\subsection{Inter-speaker Classification}
 In this experiment, we examined inter-speaker differences on the i-vector system. That is, for a given testing segment, we scored it on all of the possible models (13 speakers and 5 voice qualities  for each of them). When recording voice therapy sessions or home practice, it is inevitable that there will be speech from other people (e.g., the SLP or family members). Therefore, it is important for the system to differentiate the patient's voice from other people's voices while also detecting the patient's voice quality (i.e., combined tasks. The result of this experiment is summarize in TABLE. \ref{intereer}. As expected, the performance of the system is worse because a portion of the testing clips are classified as other actors. Although the system performs worse in combined tasks, the highest accuracy achieved (97.1\%) is still sufficient for practical use.
\begin{table}[!htb]
\renewcommand*{\arraystretch}{1.2}
\caption{Accuracy of inter-speaker experiment of two classifiers}
\label{intereer}
\centering
\begin{tabular}{lcccccc}
\hline\hline
          & \multicolumn{3}{c|}{PLDA Scoring}     & \multicolumn{3}{c}{Multi-class SVM} \\ \cline{2-7} 
          & \multicolumn{1}{c}{8s} & \multicolumn{1}{c}{4s} & \multicolumn{1}{c|}{2s} & \multicolumn{1}{c}{8s} & \multicolumn{1}{c}{4s} & \multicolumn{1}{c}{2s}  \\ \cline{2-7}
Average & \textbf{97.1\%}                           & \multicolumn{1}{c}{94.4\%} & \multicolumn{1}{c|}{89.9\%} & \textbf{94.6\%}                           & \multicolumn{1}{c}{91.8\%} & \multicolumn{1}{c|}{87.4\%} \\ \hline\hline
\end{tabular}
\end{table}


\section{discussion}
\label{sec:discussion}
From the experiment we can see that i-vector system is more accurate, it can detect more subtle voice qualities than the traditional acoustic measure system. The i-vector based system exceeded the system built on traditional measures in both intra-speaker and inter-speaker voice quality identification. This may be because, when individuals vary their voice quality, they vary many other aspects of speech production that are not captured by traditional voice measures only. 



In this study, vocal qualities were separated  by instructing participants to speak in a consistent voice quality. For example, in the ‘fry” quality, participants attempted to use fry in every word of each sentence, whereas normal speakers will typically do so only in final words of a phrase. Thus, detection of problematic vocal qualities may be more challenging when they are intermittently produced. Thus, future work will include development of an approach to scale the particular mode after its classification. For example, the system may classify a voice as “23\% fry” rather than  clarifying the entire segment as “fry.”  Furthermore, the present system may be used to automatically analyze patient home practice or mobile ambulatory recordings. In this manner, voice qualities can be quantified without requiring the clinician or investigator’s perceptual analysis of all recordings: a time-consuming process. Since patients are likely to use these to a degree rather than consistently holds utility for automatic analysis of extra-clinical patient practice recordings for study of patient adherence to voice therapy. Furthermore, the target voice in a voice patient may only be subtly different from their habitual voice. So in the end we will need a system that can classify voices that are only subtly  different,and based on even shorter segments.  So for data collection with patients they don’t need to be in the target voice for 2 minutes straight, but just in 20 sec chunks of good talking, and then coach and then record again, which will make the system easier to use. To build a practical application, We also need to analyze the performance of the system under noisy condition. MFCC features is vulnerable to noise and the performance might degrade when we have a low Signal-to-Noise Ratio (SNR) utterances. Augmentation with noise might be necessary  when turning the system into a practical application.


\section{Conclusion}
\label{sec:conclusion}
We have demonstrated that it is possible to classify dramatically different intra-speaker voice qualities, including normal, breathy, fry, twang, and hyponasal, using i-vectors. This successful classification provides proof of concept for classifying voice differences within clinical patients . The next step may be to train the system on more subtle voice quality differences, and subsequently, to detect the habitual voice quality and treatment target in actual voice patients. This work may ultimately  lead to the the production of feedback mechanisms for extra-clinical patient voice use, as the system is able to detect whether a person is using their normal habitual voice and a variety of qualities that may represents problematic vocal use or conversely, vocal targets. This shows the potential of this system to build into an actual product for voice therapy.

\bibliography{main}
\end{document}